\documentclass[runningheads]{llncs}
\usepackage[T1]{fontenc}
\usepackage{graphicx} 
\usepackage{setspace}
\usepackage{soul}
\usepackage{amssymb}
\usepackage{amsmath}
\usepackage{pifont}
\usepackage{tikz}
\usepackage{tikz, forest}
\usepackage{pgfplots, pgfplotstable}
\pgfplotsset{width=10cm,compat=1.9}
\usepackage{filecontents}
\usepackage{multirow}
\usepackage{mathtools}
\usepackage{booktabs}
\usepackage{xcolor}
\usepackage{rotating} 
\usepackage{url}
\usepackage{hyperref}
\usepackage{array}   
\usepackage[separate-uncertainty]{siunitx}
\usepackage{etoolbox}
\usepackage{caption}
\usepackage{floatrow}
\usepackage{tablefootnote}
\renewrobustcmd{\bfseries}{\fontseries{b}\selectfont}
\newcolumntype{C}{\num{S}} 

\definecolor{cbBlack}{HTML}{000000}
\definecolor{cbOrange}{HTML}{E69F00}
\definecolor{cbBlue}{HTML}{56B4E9}
\definecolor{cbGreen}{HTML}{009E73}

\definecolor{cbTol1}{HTML}{DDCC77}
\definecolor{cbTol2}{HTML}{CC6677}
\definecolor{cbTol3}{HTML}{AA4499}
\definecolor{cbTol4}{HTML}{882255}

\usepgfplotslibrary{fillbetween}
\usetikzlibrary{arrows.meta}
\usetikzlibrary{positioning}
\usetikzlibrary{shapes.geometric}   

\newcommand{\img}{\mathcal{I}}

\newcommand{\bx}{\boldsymbol{x}}
\newcommand{\bz}{\boldsymbol{z}}
\newcommand{\bmu}{\boldsymbol{\mu}}
\newcommand{\bb}{\boldsymbol{b}}

\newcommand{\apr}{\textsc{APR}}
\newcommand{\mpr}{\textsc{MPR}}
\newcommand{\mar}{\textsc{MAR}}
\newcommand{\fscore}{$F_1$}

\renewcommand{\L}{\mathcal{L}}

\newcommand{\ntestpat}{438}
\newcommand{\ntestvert}{1743}

\newcommand{\xmark}{\text{\ding{55}}}
\newcommand{\cmark}{\text{\ding{51}}}

\usepackage{subfig}

\title{Explainable vertebral fracture analysis with uncertainty estimation using differentiable rule-based classification}

\begin{document}

\author{Victor Wåhlstrand Skärström\inst{1}\orcidID{0000-0001-6569-120X} \and
Lisa Johansson\inst{2}\orcidID{0009-0005-6005-3838} \and
Jennifer Alvén\inst{1}\orcidID{0000-0003-4195-9325} \and
Mattias Lorentzon\inst{2}\orcidID{0000-0003-0749-1431}  \and
Ida Häggström\inst{1,2}\orcidID{0000-0001-9178-6683}}

\authorrunning{V. Wåhlstrand Skärström et al.}

%
\institute{
    Chalmers University of Technology, SE-41296 Göteborg, Sweden \and
    Sahlgrenska Academy at University of Gothenburg, SE-40530 Göteborg, Sweden
}

\maketitle

\section*{Abstract}
We present a novel method for explainable vertebral fracture assessment (XVFA) in low-dose radiographs using deep neural networks, incorporating vertebra detection and keypoint localization with uncertainty estimates. We incorporate Genant's semi-quantitative criteria as a differentiable rule-based means of classifying both vertebra fracture grade and morphology. Unlike previous work, XVFA provides explainable classifications relatable to current clinical methodology, as well as uncertainty estimations, while at the same time surpassing  state-of-the art methods with a vertebra-level sensitivity of 93\% and end-to-end AUC of 97\% in a challenging setting. Moreover, we compare intra-reader agreement with model uncertainty estimates, with model reliability on par with human annotators.

\section{Introduction}
Vertebral compression fractures, osteoporotic deformations due to the loss of bone mass in the spine, are a common and serious complication of osteoporosis and a leading cause of impaired mobility and increased mortality \cite{Lorentzon2019}. Vertebral fracture assessment (VFA) using X-ray scans is a critical screening method used to detect osteoporotic fractures and apply interventive measures early \cite{Johansson2020}. The number and location of fractures is an important indicator of disease progression, but VFA is challenging, requiring time-consuming annotation of vertebrae by expert clinicians for reliable diagnoses. Patients are often elderly with difficulty holding position, and scans are commonly low-dose, decreasing image quality.

\begin{figure}[t!]
    \centering
    \resizebox{!}{2.3cm}{
        \input{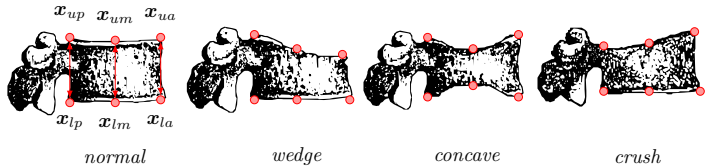}
    }
    \caption{\textit{Visualization of GSQ.} Vertebral fracture morphology in Genant's system, with landmarks $\bx$ (dots). The posterior, middle, and anterior heights (lines) are used to determine deformation morphology and severity. Fig. adapted from \cite{Genant1993}.}
    \label{fig:vertebral_height}
\end{figure}

Genant's semi-quantitative method (GSQ) is a common means of VFA, where vertebrae in a spinal X-ray scan are annotated with six keypoints, and fractures are classified per morphology and severity \cite{Genant1993}, see Figure~\ref{fig:vertebral_height}. In the semi-quantitative approach, readers make a compound judgement based on keypoint positions, visual features and adjacent vertebrae. The complexity of VFA makes the adoption of automated methods slow, and it is inhibited by poor understanding by clinicians, and low reliability and sensitivity of existing methods.

Previous work on automated VFA has focused on image level classification of the presence of vertebrae, due to the expensive nature of annotation. However, research indicates that the number and severity of compressions is highly predictive of future fractures.\cite{lindsay2001} Most authors employ convolutional neural networks (CNNs) for grade classification \cite{murata2020artificial,Dong2022,Monchka2022}, using Grad-CAM \cite{Selvaraju2016} for weak localization and model interpretability. None of these methods provide direct localization of vertebrae, and methods like CAM and saliency maps do not explain model reasoning, only visually ranking feature contributions to predictions without providing tractable reasoning on their use in the model. Deep learning methods are thus commonly criticized for their ``black box'' classification and poor explainability, reducing trust and inhibiting the adaptation of methods in healthcare. Various approaches have addressed interpretability post-hoc for neural networks, as surveyed by e.g. Kouchaki et al. \cite{Kouchaki2023}, while \textit{interpretability} may be defined as an understanding of how an underlying model works, \textit{explainability} reflects how the model came up with a given result. \cite{ISO_explainability}  Unlike neural networks, decision trees are intrinsically explainable, but typically not differentiable.

The localization of vertebrae has been studied in various imaging modalities, e.g. planar X-ray and computed tomography, e.g. using CNNs and graph neural networks \cite{burgin2023,Wu2023}. For VFA, localization is not enough, and lack of annotated vertebrae make detection approaches rare. Recent end-to-end contributions include Cheng et al. \cite{Cheng2024}, presenting a method of classification of trauma-induced fractures, where segmented vertebra features such as ridge heights were used as input to a random forest, yielding high sensitivity. Like previous work, their method operates on fairly high-contrast radiographs and does not differentiate fracture severity or relate their classifications to existing clinical methodology. Beside explainability, previous work rarely address model trust, an important factor in high stakes applications such as healthcare. An aspect of this is model uncertainty, for example confidence or error intervals imposing limits on model applicability. Dong et al. \cite{Dong2023} use CNN ensembling for prediction, showing reliability across datasets. Residual log-likelihood \cite{li_2021} introduces a fast and light-weight alternative to expensive ensembling for regression tasks, using a normalizing flow \cite{Rezende2015} to model the error distribution of regressed points.

\subsection{Contributions}
To our knowledge, no previous work has addressed VFA in low-dose, high-noise radiographs by incorporating GSQ directly into neural network classification. Our main contributions are creating a novel, explainable and reliable detection model of vertebral fractures through \textit{i}) state-of-the-art detectors for vertebra localization, \textit{ii}) estimation of vertebra keypoints with uncertainty quantification and \textit{iii}) an innovative classification scheme using estimated vertebral keypoints to compute rule-based objective from GSQ. For this purpose we implement a differentiable decision tree, which enables us to propagate the keypoint uncertainty to the final fracture classification.

\section{Methodology}
\begin{figure}[t!]
    \centering
    \resizebox{\textwidth}{!}{
        \input{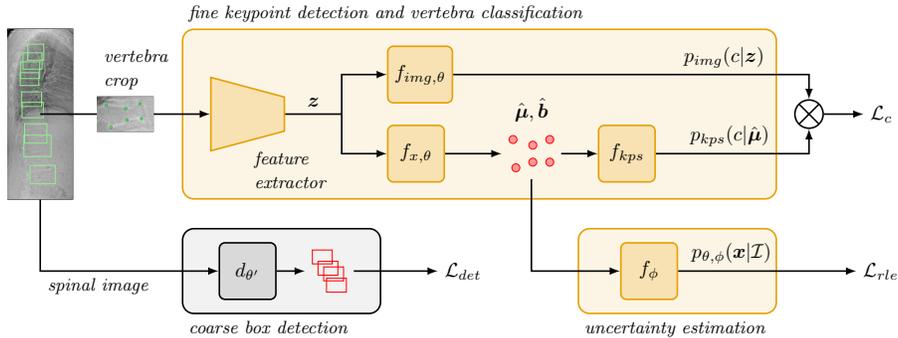}        
    }
    \caption{\textit{Model overview.} A detector $d_{\theta'}$ is trained to estimate bounding boxes (\textit{red}) from vertebra in a spinal image (\textit{green}). Ground truth crops with keypoints (\textit{green}) are separately used to train a keypoint regressor $f_{x,\theta}$, subsequent vertebra classifiers $f_{img, \theta}, f_{kps}$, as well as estimate the keypoint distribution (\textit{red points}).}
    \label{fig:architecture}
\end{figure}

An overview of XVFA is shown in Figure~\ref{fig:architecture}. We employ a two-stage method of coarse bounding box detection of vertebrae using a state-of-the-art (SOTA) detector, and regress vertebral keypoints and estimate of the empirical model error distribution using normalizing flows. Vertebrae are subsequently classified using a combination of GSQ criteria and a neural image classifier.

\subsection{Coarse bounding box detection}
In this paper, we use a SOTA detection transformer architecture, DETR with Improved DeNoising Anchor Boxes (DINO) with a ResNet50 backbone for localization \cite{zhang2022dino,He2015}, having lower footprint and higher accuracy than its progenitor, DETR \cite{Carion2020}. We use the same loss $\L_{det}$ as defined in DINO and DETR, 
\begin{equation}
    \label{eq:dino}
    \L_{det} = \sum_i-\log{\hat{p}}(c_i) + \textbf{I}_{c_i\neq\emptyset}\left( \lambda_{iou}\L_{iou}(b_i,\hat{b}_i) + \lambda_{\ell1}||b_i-\hat{b}_i||_1 \right),
\end{equation}
computed between ground truth bounding boxes $b$ and predicted boxes $\hat{b}$ matched with the Hungarian algorithm and regressed with generalized IoU loss $\L_{iou}$, \cite{Rezatofighi2019} and $\ell1$-loss. $\hat{p}(c)$ denotes the predicted class of the matched predicted object. If the matched object is predicted to belong to a non-object class, it is omitted from the loss, denoted with the indicator function, $\textbf{I}_{c_i\neq\emptyset}$. In our case, we predict the presence or absence of vertebrae, corresponding to two classes in DETR. 

\subsection{Fine vertebra detection}
\subsubsection*{Genant's Semi-Quantitative Method}
Consider vertebra keypoints $\bx_{ij}=(x_{ij},y_{ij})$, as defined in Figure~\ref{fig:vertebral_height}. 
The morphology of fractures varies between \textit{normal}, \textit{concave}, \textit{wedge} and \textit{crush} deformities, quantified in terms of vertebral heights $h$ between the posterior, middle or anterior parts of the ridges:
\begin{equation}
\label{eq:heights}
    h_p = || \bx_{up} - \bx_{lp} ||,\quad h_m = || \bx_{um} - \bx_{lm} ||,\quad h_a = || \bx_{ua} - \bx_{la} ||, 
\end{equation}
where $||\cdot||$ denotes the 2-norm. Fracture morphology may be characterized through the anterior-posterior, middle-posterior and middle-anterior ratios between heights,
\begin{equation}
    \text{APR} = \frac{h_a}{h_p},\quad \text{MPR} = \frac{h_m}{h_p},\quad \text{MAR} = \frac{h_m}{h_a}.
\end{equation}
A normal vertebra is expected to be approximately rectangular and have equal height ratios of 1, within a tolerance $20\%$. A wedge deformity has $\mpr<1$ and $\mar>1$, and inversely for crush fractures. Concave fractures have $\mpr<1$ and $\mar<1$. The inverse case of convex vertebrae are considered normal.

According to GSQ, a mild fracture is defined as a relative ratio loss of 20-25\% of vertebral height, a moderate fracture as 25-40\% loss, and a severe fracture as more than 40\% loss. Due to the small differences between classes, they are typically grouped as \textit{normal+mild} and \textit{moderate+severe}.


\subsubsection*{Keypoint estimation using residual log-likelihood}
Using a ResNet18 \cite{He2015}, vertebra images are encoded as features $\bz$. As per residual log-likelihood estimation (RLE) \cite{li_2021}, we estimate the log-likelihood $\log{p_{\theta,\phi}(\bx|\bz)}$ of vertebra keypoints $\bx$ conditioned on the visual features $\bz$ by modelling the keypoint error distribution with a normalizing flow, $f_\phi$, in our case RealNVP \cite{realnvp}.
%
%
Like in maximum likelihood estimation, we assume an approximate target distribution, e.g. Gaussian or Laplace, where a trainable keypoint regressor predicts the centre and scale $f_{\theta}(\bz) = (\hat{\boldsymbol{\mu}}, \hat{\boldsymbol{b}})$ of the distribution, minimizing the  negative log-likelihood $-\log{p_{\theta,\phi}(\bx|\bz)}$.
The normalizing flow maps an initial distribution to a flow distribution $\bx'\sim p_\phi$ such that the final distribution of keypoints $\bx|\bz\sim p_{\theta,\phi}$ is given by shifting and scaling, $\bx = \hat{\bb}\,\bx' + \hat{\bmu}$. The log-likelihood can thus be written
\begin{equation}
\label{eq:logphitheta}
    \log{p_{\theta,\phi}(\bx|\bz)} = \log p_\phi\left(\frac{\bx-\hat{\bmu}}{\hat{\bb}}\right) - \log \hat{\bb}.
\end{equation}
Specifically, we assume that the predicted normalizing flow decomposes into a base distribution $q(\bx')$ (e.g. a Gaussian) and a residual distribution $g_\phi(\bx')$ up to a constant factor, such that $p_\phi=q+g_\phi$, yielding
\begin{equation}
   \L_{rle} =  \sum_i -\log q\left(\frac{\bx_i-\hat{\bmu}_i}{\hat{\bb}_i}\right) - \log g_\phi\left(\frac{\bx_i-\hat{\bmu}_i}{\hat{\bb}_i}\right) + \log \hat{\bb}_i.
\end{equation}
For more details, please refer to the original RLE paper. We estimate the uncertainty $\boldsymbol{\varepsilon}$ for a predicted keypoint $\hat{\bmu}$ as the quantiles at a pre-set level $\alpha$ as
\begin{equation}
    \int_{\hat{\bmu}-\boldsymbol{\varepsilon}}^{\hat{\bmu}+\boldsymbol{\varepsilon}} p_{\theta, \phi}(\bx | \bz)\,\mathrm{d}\bx = \alpha.
\end{equation}




\subsubsection*{Vertebra classification}
We consider the classification of vertebrae as a separable product of keypoint features and visual features
\begin{equation}
 \label{eq:classification}
     p(c|\img, \bx) = p_{kps}(c|\bx)\,p_{img}(c|\bz).
 \end{equation}
 where $p_{img}(c|\bz)$ is the output of a single-layer neural network from the visual features $\bz$, and $p_{kps}(c|\bx)$ is a probability from a GSQ-based differentiable classification.
We construct a \textit{fuzzy} decision function classifier $f_{kps}$ by approximating the hard conditions of GSQ with differentiable counterparts. We replace operations ``greater than'' and ``less than'' with a sigmoid with a threshold $t$, and logical ``and'' with the minimum operation, as per the principles of e.g. \cite{Mamdani1974}, yielding soft class predictions $c|\bx\sim p_{kps}$ given the vertebra keypoints.
For training the classification model, we use a weighted cross-entropy loss
\begin{equation}
    \label{eq:loss_tot}
     \L_{c} = -\sum_i\sum_c w_c \mathrm{\textbf{I}}_c\left(\lambda_{kps}\log{p_{kps}(c|\bx_i)} + \lambda_{img}\log{p_{img}(c|\bz_i)}\right),
 \end{equation}
where  $\lambda_{img}, \lambda_{kps}$ are hyperparameters, $w_c$ are class weights for grade and morphology, and $\mathrm{\textbf{I}}_c$ is an indicator function for belonging to class $c$. This formulation allows us to easily separate contributions to the classification, e.g. by setting $\lambda_{img}=0$ during training, and letting $p(c|\img, \bx) = p_{kps}(c|\bx)$ during inference.





\section{Experiments and Results}
\subsection{Dataset}
The Sahlgrenska University Hospital Prospective Evaluation of Risk of Bone Fractures (SUPERB)  is a population-based study of 3028 older women aged $77.8 \pm 1.6$ from Gothenburg, Sweden \cite{Johansson2020,Lorentzon2019}.
At baseline, $13$ thoracic and lumbar vertebrae were annotated with keypoints and analyzed using lateral low-dose radiographs, and prevalent fracture grade and morphology according to GSQ by two annotators with a high degree of inter-reader agreement. Due to the noisy quality of the low-dose X-ray and the expensive nature of annotation, the number of labelled vertebrae is limited, ranging from 1 to 13 per patient. Out of the full set of patients, 2919 had digitized annotations of 11,605 vertebrae. See suppl. Table~1 for an overview of the data.

\subsection{Model implementation and training}
Models were implemented in PyTorch 2.2 
on two NVIDIA RTX A6000 GPUs. All metrics were computed using the \texttt{scikit-learn} library, v. 1.3.0 
. We randomly held out 438 out of 2919 patients (15\%) for final testing, and divided the remaining 2481 (85\%) patients into 5 folds for cross-validation for model training and tuning. Due to the low presence of crush deformities (suppl. Table~1), wedge and crush classes were combined into a \textit{wedge-like} class.

Missing labels (vertebrae) is often detrimental for detection models \cite{xu2019missing}. Here we imputed missing keypoints using $k$-nearest neighbours before training, and downweighed their contribution to eq.~\eqref{eq:dino} by a factor $10^{-3}$.
Models were trained with the Adam optimizer.
DINO was first trained using the full data with imputations, and then fine-tuned with just the fully annotated data. Full images were downsampled to half the original size and padded to a uniform size per batch, and vertebra  bounding boxes were resized to (224, 224). Standard augmentation techniques such as rotations, flipping, and jittering were applied. For complete hyperparameters chosen with grid search, see the suppl. Table~2. 


\subsection{Ablation study}
We investigated the contributions of four combinations to eq.~\eqref{eq:loss_tot}: keypoint regression ($\lambda_{img}=\lambda_{kps}=0$), GSQ-guided loss ($\lambda_{img}=0, \lambda_{kps}=1$), classification from image features ($\lambda_{img}=1, \lambda_{kps}=0$) and combining both ($\lambda_{img}=1, \lambda_{kps}=1$).
\begin{table}[t!]
\centering
\caption{\textit{Ablation study of loss components.} 5-fold comparison of the vertebra grade classification on the test set ($N_{vertebra}=\ntestvert$). \fscore, specificity and sensitivity computed at the Youden operating point. ($^\ast$) indicates significant difference to the full model using a Bonferroni $t$-test at $\alpha=0.05$.}
\label{tab:ablation}
{
\renewcommand{\arraystretch}{1}
    \begin{tabular}{c|ccc|SSSS}
    \hline

      \multicolumn{1}{c}{Classification} &
      \multicolumn{1}{c}{$\L_{rle}$} &
      \multicolumn{1}{c}{$\L_{kps}$} &
      \multicolumn{1}{c}{$\L_{img}$} &

      \multicolumn{1}{c}{AUC} &
      \multicolumn{1}{c}{\fscore} &
      \multicolumn{1}{c}{sensitivity} &
      \multicolumn{1}{c}{specificity} \\

\hline
\multirow{4}{*}{\shortstack[c]{\textit{normal}\\ \textit{vs.}\\ \textit{rest}}} &
 \cmark & & 
 & \num{0.96+-0.01}$^\ast$
 & \num{0.51+-0.03}$^\ast$
 & \num{0.90+-0.03}
 & \num{0.90+-0.01}$^\ast$
\\
& \cmark & \cmark &    
& \num{0.92+-0.01}$^\ast$
& \bfseries\num{0.90+-0.02 }$^\ast$
& \num{0.81+-0.04 }$^\ast$
& \num{0.90+-0.06}
\\
& \cmark & & \cmark 
& \num{0.98+-0.01}
& \num{0.70+-0.10}
& \num{0.90+-0.04}$^\ast$
& \bfseries\num{0.96+-0.02}
\\
& \cmark & \cmark & \cmark 
& \bfseries\num{0.98+-0.01}
& \num{0.67+-0.07}
& \bfseries\num{0.92+-0.04}
& \num{0.95+-0.02}
\\
\hline
\multirow{4}{*}{\shortstack[c]{\textit{normal+}\\\textit{mild}\\ \textit{vs.}\\ \textit{moderate+}\\\textit{severe}}} &
 \cmark & & 
 & \num{0.97+-0.01}$^\ast$
 & \num{0.42+-0.05}$^\ast$
 & \num{0.93+-0.03}
 & \num{0.90+-0.03}$^\ast$
 \\
& \cmark & \cmark & 
& \num{0.97+-0.01}$^\ast$
& \num{0.45+-0.08}$^\ast$
& \num{0.91+-0.03}
& \num{0.91+-0.03}$^\ast$
\\
& \cmark & & \cmark 
& \num{0.98+-0.01}
& \num{0.71+-0.09}
& \num{0.928+-0.032}
& \num{0.97+-0.01}
\\
& \cmark & \cmark & \cmark 
& \bfseries\num{0.99+-0.01}
& \bfseries\num{0.73+-0.08}
& \bfseries\num{0.934+-0.039}
& \bfseries\num{0.97+-0.01}
\\
\hline
  \end{tabular}
  }
\end{table}
Results from the ablation study are shown in Table~\ref{tab:ablation} calculated as mean and standard deviation on the test set over the 5 models from each fold. AUC, \fscore, sensitivity and specificity are computed as one-vs.-rest binary classification for \textit{normal} vertebrae vs. fractures and \textit{normal+mild} vs. \textit{moderate+severe}. See the supplementary for results on morphology and confusion matrices.

Our proposed method, using both the rule-based loss and image feature loss outperforms the rest on identifying the clinically most relevant grouping of \textit{normal+mild} vertebrae, with an AUC of $0.99$. While using the image-based loss yields competitive results on its own, it retains the issues of explainability and lacks connection to GSQ. Fracture predictions may be visualized in terms of decision borders for \mpr~and \mar, see  Figure~\ref{fig:marmpr_type}-\subref*{fig:marmpr_grade}, showing that the classification of individual vertebrae in the test set closely align with GSQ, providing a useful tool to find disagreements between image and keypoint-based classifications.

\subsection{SOTA model comparison}
In the end-to-end results, we compared results on vertebra level and on patient-aggregated level using our best performing model ablation. We implemented a similar CNN model to \cite{murata2020artificial,Monchka2022}, using a pre-trained Inception-ResNet-v2, classifying max severity grade in the image (learning rate of $10^{-5}$, batch size 16), comparing with XVFA by aggregating per patient post-classification. As a baseline for vertebra-level detection, we trained DINO models with the same settings as earlier, but classifying grade. We compared with \cite{Dong2022} (equiv. to our method with $\lambda_{kps}=0$) and \cite{Cheng2024}, using a random forest ($100$ trees, depth $2$) on vertebra features $h_a, h_m, h_p, \apr, \mpr, \mar$. The results are shown in Table~\ref{tab:baselines}.

XVFA outperforms the SOTA and DINO baseline on the important screening metrics, AUC and sensitivity, and achieves comparable results on the rest. Like in the ablation study, using a convolutional image classifier is competitive, but still lacks the explicit relation to GSQ. It is notable that the approach of \cite{murata2020artificial,Monchka2022} vastly underperforms on our images, likely due to the low signal-to-noise ratio of the low-dose radiograph images. Lastly, as demonstrated by this paper, decision trees can solve the fracture classification task well, and it is no surprise that the random forest approach has competitive results. However, as is also the case with purely image-based classification, random forests are not explainable and directly relatable to clinical methodology.

\begin{table}[t!]
\centering
    \caption{\textit{SOTA comparison.} 5-fold comparison of SOTA methods on test set on vertebra detection and classification (\textit{vert.}, $N_{vertebra}=\ntestvert$) and patient/image-level classification (\textit{pat.}, $N_{patient}=\ntestpat$) for \textit{normal} vertebrae+\textit{mild} fractures vs. \textit{moderate}+\textit{severe} fractures. ($^\ast$) indicates significant difference to our model using a Bonferroni $t$-test at $\alpha=0.05$.}
    \label{tab:baselines}
{
\renewcommand{\arraystretch}{1}
    \begin{tabular}{>{\scriptsize}cccc|SSSS}
    \hline

      \multicolumn{1}{c}{Method} &
      \multicolumn{1}{c}{Level} &
      \multicolumn{1}{c}{Ref.} &
      \multicolumn{1}{c}{Expl.} &

      \multicolumn{1}{c}{AUC} &
      \multicolumn{1}{c}{\fscore} &
      \multicolumn{1}{c}{sensitivity} &
      \multicolumn{1}{c}{specificity} \\

\hline

DINO & vert. & - & \text{\xmark}
& \num{0.95+-0.02}
& \num{0.33+-0.06}$^\ast$
& \num{0.90+-0.05}
& \num{0.85+-0.06}$^\ast$
\\

 CNN & vert. & \cite{Dong2022} & \text{\xmark}
& \num{0.97+-0.01}
& \bfseries\num{0.72+-0.04}
& \num{0.90+-0.04}
& \bfseries\num{0.976+-0.004}
\\
 RF & vert. & \cite{Cheng2024} & \text{\xmark}
& \num{0.95+-0.02}
& \num{0.56+-0.11}
& \num{0.85+-0.03}
& \num{0.95+-0.03}
\\

 XVFA\,(\textit{ours}) & vert. & - & \text{\cmark}
& \bfseries\num{0.97+-0.02}
& \num{0.65+-0.11}
& \bfseries\num{0.90+-0.06}
& \num{0.96+-0.02}
\\
\hline
\hline
 CNN & pat. & \cite{murata2020artificial,Monchka2022} & \text{\xmark}
 & \num{0.81+-0.09}
 & \num{0.62+-0.11}
 & \num{0.60+-0.16}$^\ast$
 & \bfseries\num{0.90+-0.04}
 \\
 XVFA\,(\textit{ours}) & pat. & - & \text{\cmark}
& \text{-}
& \bfseries\num{0.70+-0.10}
& \bfseries\num{0.92+-0.05}
& \num{0.86+-0.08}
\\
\hline
  \end{tabular}
  }
\end{table}

\subsection{Uncertainty quantification}
\begin{figure}[t!]
     \centering
    \sidesubfloat[]{
      \input{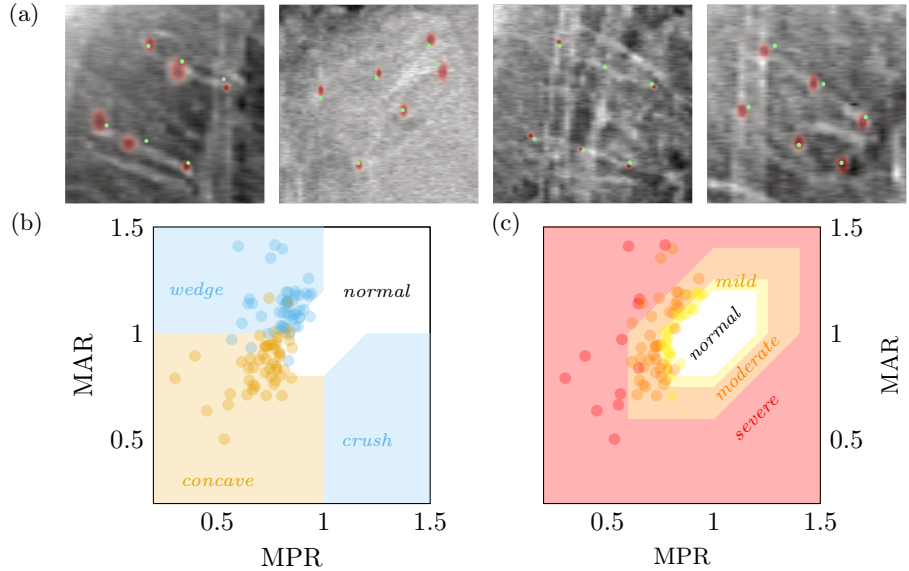}
        \label{fig:uncertainty}
    }\qquad\newline
    \sidesubfloat[]{
            \begin{tikzpicture}
    \begin{axis}[
        xlabel=MPR,
        ylabel=MAR,
        xmin=0.2,
        xmax=1.5,
        ymin=0.2,
        ymax=1.5,
        width=0.5\textwidth,
        axis equal image,
        ytick style={draw=none},
        xtick style={draw=none}
    ]

    \addplot+[mark=none, name path=x1, opacity=0] coordinates {(0,0) (0,1)};
    \addplot+[mark=none, name path=x2, opacity=0] coordinates {(0,1) (0,3)};
    
    \addplot+[mark=none, name path=y1, opacity=0] coordinates {(0,0) (0.8,0)};
    \addplot+[mark=none, name path=y2, opacity=0] coordinates {(0.79,0) (1.0,0)};
    \addplot+[mark=none, name path=y3, opacity=0] coordinates {(1,0) (1.2,0)};
    \addplot+[mark=none, name path=y4, opacity=0] coordinates {(1.19,0) (2.1,0)};

    \addplot+[mark=none, name path=ytop1, opacity=0] coordinates {(0,2.1) (0.8,2.1)};
    \addplot+[mark=none, name path=ytop2, opacity=0] coordinates {(0.79,2.1) (1,2.1)};

    \addplot+[mark=none, black, solid, name path=A, opacity=0] coordinates {(0,1) (0.8,1)};
    \addplot+[mark=none,  black, solid, name path=B, opacity=0] coordinates {(1.2,1) (2.1,1)};

    \addplot+[mark=none,  black, solid, name path=C, opacity=0] coordinates {(1,1.2) (1,2.1)};
    \addplot+[mark=none,  black, dashed, name path=C, opacity=0] coordinates {(1,0) (1,0.8)};
    \addplot+[mark=none,  black, dashed, name path=D, opacity=0] coordinates {(1,1.2) (1,2.1)};
    
    \addplot+[mark=none,  black, solid, name path=E, opacity=0] coordinates {(0.8, 0.8) (0.8, 1)};
    \addplot+[mark=none,  black, solid, name path=F, opacity=0] coordinates {(0.8, 0.8) (1, 0.8)};
    \addplot+[mark=none,  black, solid, name path=G, opacity=0] coordinates {(1, 0.8) (1.2, 1)};
    \addplot+[mark=none,  black, solid, name path=H, opacity=0] coordinates {(0.8, 1) (1, 1.2)};        

   \addplot[cbOrange!20] fill between[of=A and y1];
   \addplot[cbOrange!20] fill between[of=F and y2];

   \addplot[cbBlue!20] fill between[of=A and ytop1];
   \addplot[cbBlue!20] fill between[of=H and ytop2];
   
   \addplot[cbBlue!20] fill between[of=B and y4];
   \addplot[cbBlue!20] fill between[of=G and y3];

    \addplot+[
        scatter, 
        only marks,  
        point meta=explicit symbolic,
        scatter/classes={
            0={mark=o, color=cbBlack, opacity=0.0},
            1={mark=*, color=cbBlue, opacity=0.3},
            2={mark=*, color=cbOrange, opacity=0.3},
            3={mark=*, color=cbGreen, opacity=0.3}
            },
        ] table [
            x=mpr, 
            y=mar, 
            meta=type,
            col sep=comma, 
        ] {figures/data/vertebra_params.csv};

    \node at (axis cs: 0.4, 1.2) {\scriptsize\textit{\textcolor{cbBlue}{wedge}}};
    \node at (axis cs: 1.2, 0.5) {\scriptsize\textit{\textcolor{cbBlue}{crush}}};
    \node at (axis cs: 0.5, 0.3) {\scriptsize\textit{\textcolor{cbOrange}{concave}}};
    \node at (axis cs: 1.25, 1.2) {\scriptsize\textit{normal}};

    
    \addplot+[mark=none,  black, solid, name path=border, opacity=1] coordinates {
        (0.2, 0.2) 
        (1.5, 0.2)
        (1.5, 1.5)
        (0.2, 1.5)
        (0.2, 0.2)
        };


    \end{axis}
\end{tikzpicture}
        \label{fig:marmpr_type}
    }~
    \sidesubfloat[]{
            \begin{tikzpicture}[
    dot/.style = {circle, fill, minimum size=#1,
                  inner sep=0pt, outer sep=0pt, fill=red!40,
            draw=red,},
]
    \begin{axis}[
        xlabel={\footnotesize MPR},
        ylabel={\footnotesize MAR},
        xmin=0.2,
       xmax=1.5,
        ymin=0.2,
        ymax=1.5,
        width=0.5\textwidth,
        axis equal image,
        ylabel near ticks, yticklabel pos=right,
        ytick style={draw=none},
        xtick style={draw=none}
    ]

    \addplot+[mark=none, name path=x1, opacity=0] coordinates {(0,0) (0,1)};
    \addplot+[mark=none, name path=x2, opacity=0] coordinates {(0,1) (0,3)};
    
    \addplot+[mark=none, name path=y1, opacity=0] coordinates {(0,0) (0.8,0)};
    \addplot+[mark=none, name path=y2, opacity=0] coordinates {(0.79,0) (1.0,0)};
    \addplot+[mark=none, name path=y3, opacity=0] coordinates {(1,0) (1.2,0)};
    \addplot+[mark=none, name path=y4, opacity=0] coordinates {(1.19,0) (2.1,0)};

    \addplot+[mark=none, name path=ytop1, opacity=0] coordinates {(0,2.1) (0.8,2.1)};
    \addplot+[mark=none, name path=ytop2, opacity=0] coordinates {(0.79,2.1) (1,2.1)};

    \addplot+[mark=none, black, dashed, name path=A, opacity=0] coordinates {(0,1) (0.8,1)};

     \addplot+[mark=none,  black, solid, name path=border, opacity=1] coordinates {
        (0.2, 0.2) 
        (1.5, 0.2)
        (1.5, 1.5)
        (0.2, 1.5)
        (0.2, 0.2)
        };

    \addplot+[mark=none,  black, solid, name path=healthy, opacity=0] coordinates {
        (0.8, 0.8) 
        (0.8, 1)
        (1, 1.2)
        (1.2, 1.2)
        (1.2, 1)
        (1, 0.8)
        (0.8, 0.8)
        };

    \addplot+[mark=none,  yellow!30, solid, name path=mild, opacity=1] coordinates {
        (0.751, 0.751) 
        (0.751, 1)
        (1, 1.251)
        (1.251, 1.251)
        (1.251, 1)
        (1, 0.751)
        (0.751, 0.751)
        };

     \addplot+[mark=none,  orange!30, solid, name path=moderate, opacity=1] coordinates {
        (0.6, 0.6) 
        (0.6, 1)
        (1, 1.4)
        (1.4, 1.4)
        (1.4, 1)
        (1, 0.6)
        (0.6, 0.6)
        };
        
     \addplot+[mark=none,  red!30, solid, name path=severe, opacity=1] coordinates {
        (0.0, 0.0) 
        (0.0, 2.9)
        (2.1, 2.1)
        (2.1, 0)
        (0,0)
        };
        \addplot[yellow!30] fill between[of=healthy and mild];
        \addplot[orange!30] fill between[of=mild and moderate];
        \addplot[red!30] fill between[of=moderate and severe];

    \addplot+[
        scatter, 
        only marks,  
        point meta=explicit symbolic,
        scatter/classes={
            0={mark=*, color=cbBlack, opacity=0.0},
            1={mark=*, color=yellow, opacity=0.3},
            2={mark=*, color=orange, opacity=0.3},
            3={mark=*, color=red, opacity=0.3}
            },
        ] table [
            x=mpr, 
            y=mar, 
            meta=pred_grade,
            col sep=comma, 
        ] {figures/data/vertebra_params.csv};



    \node[rotate=45] at (axis cs: 1, 1) {\scriptsize\textit{normal}};
    \node[rotate=45] at (axis cs: 1.15, 0.84) {\scriptsize\textit{\textcolor{orange}{moderate}}};
    \node[rotate=45] at (axis cs: 1.2, 0.6) {\scriptsize\textit{\textcolor{red}{severe}}};
    
    \node[rotate=0] at (axis cs: 1.1, 1.25) {\scriptsize\textit{\textcolor{cbOrange}{mild}}};


    \end{axis}
\end{tikzpicture}
   \label{fig:marmpr_grade}
  }%
        \caption{\textit{Visualizations.} (a) Test sample model prediction likelihoods (\textit{red}) and ground truth keypoints (\textit{green)}. GSQ decision borders for (b) morphology and (c) grade, for visual explanation of model results, each point corresponding to a predicted fractured vertebra in the test set.}
        \label{fig:viz}
\end{figure}

We quantified the uncertainty of individual models using the full likelihood from the normalizing flow. For the best model on the validation set, we sample 1000 keypoints with replacement from the densities, see e.g. Figure~\ref{fig:uncertainty} acquiring a mean$\pm$standard deviation for the model on the test set: AUC $0.985\pm0.002$, sensitivity $ 0.941\pm0.004$, specificity $0.988\pm0.004$, and \fscore\, of $0.842\pm0.029$. Lastly, we tested model alignment with human annotators by asking the annotator to re-analyze a random subset of 49 scans from the test set. 
The annotator had a mean intra-reader deviation of approximately $\pm4.5$ pixels, and our model had an uncertainty of $\pm5.3$ pixels within the 5th and 95th percentiles, indicating model reliability on par with human observers.
\floatsetup[figure]{style=plain,subcapbesideposition=top}




\section{Discussion and Conclusion}
We have shown that XVFA can accurately classify vertebral fractures on par with or better than SOTA, with a per-vertebra AUC of 99\% and sensitivity of 93\% and end-to-end AUC of 97\%, without sacrificing explainability. Proper uncertainty quantification yields reliability measures both in keypoint localization and fracture classification. We also demonstrate that use of black box neural classifiers is not necessarily superior to using an explainable alternative based on GSQ. Our method is also flexible, and may be updated with any SOTA detector. Moreover, while this paper has mostly focused on the established task of fracture grade classification, excellent results on morphology classification are available in the supplementary which lack comparative baselines.

However, our model does have some limitations. It does not account for additional features, such as adjacent vertebrae into classification (see e.g. \cite{Cheng2024}) and has no native means of handling missing vertebrae in annotation. A more explicit modelling of spinal structure would likely address these issues. Moreover, our approach does not distinguish \textit{epistemic} and \textit{aleatoric} uncertainty, focusing only on the latter. Normalizing flows may also have an issue with overfitting \cite{kirichenko2020normalizing}, which we did not notice in this paper, requiring research on other cohorts to validate. In spite of these shortcomings, our model displays sufficient sensitivity for screening ($> 90\%$) of fracture grades, and may provide trustworthy and explainable decision support useful in clinical work that can aid the evaluation of VFA, alleviate physician workload, and ultimately improve patient outcome.

\section{Data Use Declaration and Acknowledgment}

\noindent
The collection of data for the SUPERB study was approved by the 
Ethical Review Authority Swedish Ethical Review Authority (DNR 929-12). 
The data is not publicly available as it contains sensitive health information. The source code and trained models can be downloaded for academic or non-commercial purposes.\footnote{\url{https://github.com/waahlstrand/xvfa}}
{\bf\ackname} This study was funded in part through the AIDA project grant DNR 2021-01420.
{\bf\discintname} Outside of this research, ML has received lecture fees from Amgen, Astellas, Lilly, Meda, Renapharma,
UCB Pharma, and consulting fees from Amgen, Radius Health, UCB Pharma, Parexel International, Renapharma and Consilient Health, LJ has received lecture fees from UCB Pharma. All other authors declare no competing interests.
%
%
%
\newpage

\bibliographystyle{splncs04}
\bibliography{bibliography}

\section*{Supplementary}
\begin{table}[h!]
    \centering
    \caption{\textit{Demographic statistics.} Fractures and relevant properties of the annotated subset of the SUPERB cohort. }
    \label{tab:demographics}
    {
    \renewcommand{\arraystretch}{1}
    \begin{tabular}{c|c|c|c|ccc|ccc}
      \hline
      \multicolumn{1}{c|}{\,} &
      \multirow{2}{2cm}{\centering number of patients} &
      \multirow{2}{2cm}{\centering number of vertebrae} &
      \multirow{2}{*}{normal} &
      \multicolumn{3}{c|}{grade} &
      \multicolumn{3}{c}{morphology} \\
      
        & & &  & mild & moderate & severe & wedge & crush & concave  \\
      \hline
        \textit{train}   & 2481  & 9862 & 9319 & 227 & 222   & 94 & 309  & 229 & 5  \\
        \textit{test}    & 438   & 1743 & 1643 & 33  & 47    & 20 & 47   & 2 & 51  \\
        \cline{1-10}
        \textit{total}   & 2919  & 11605 & 10962 & 260  & 269 & 114 & 356 & 231 & 56  \\
    \hline
    \end{tabular}
    }
\end{table}


\floatsetup[figure]{style=plain,subcapbesideposition=top}
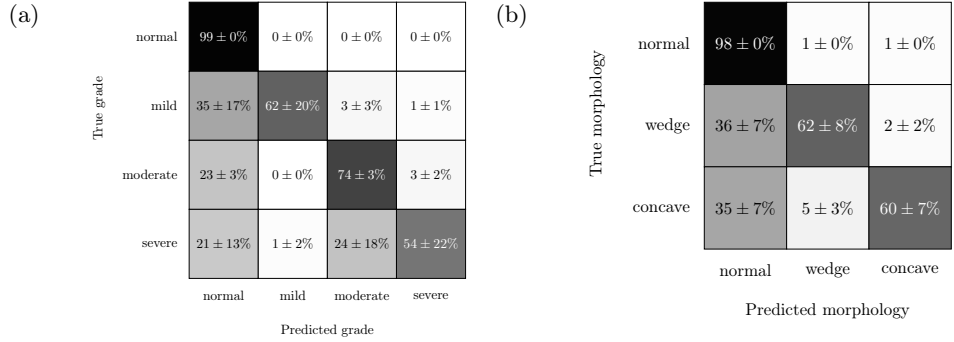
\begin{figure}[h!]
     \centering
    \sidesubfloat[]{
     \resizebox{0.45\textwidth}{!}{
\def\myConfMat{{
{1633.4, 4.0, 5.4,0.2},  
{11.4, 20.4, 1.0, 0.2},  
{10.6, 0.0, 35.0, 1.4},  
{4.2, 0.2, 4.8, 10.8}  
}}

\def\stdMat{{
{6.107373,	2.549510,	5.272571,	0.447214},  
{5.549775,	6.655825,	1.000000,	0.447214},  
{1.341641,	0.000000,	1.224745,	0.894427},  
{2.588436,	0.447214,	3.563706,	4.494441}  
}}

\def\classNames{{"normal","mild","moderate","severe"}} 

\def\numClasses{4} 

\def\myScale{1.8} 
\begin{tikzpicture}[
    scale = \myScale,
    ]

\tikzset{vertical label/.style={rotate=90,anchor=east}}   
\tikzset{diagonal label/.style={rotate=45,anchor=north east}}

\foreach \y in {1,...,\numClasses} 
{
    \node [anchor=east] at (0.4,-\y) {\pgfmathparse{\classNames[\y-1]}\pgfmathresult}; 

    \def\totSamples{0}
    \foreach \ll in {1,...,\numClasses}
    {
        \pgfmathparse{\myConfMat[\y-1][\ll-1]}   
        \xdef\totSamples{\totSamples+\pgfmathresult} 
    }
    \pgfmathparse{\totSamples} \xdef\totSamples{\pgfmathresult}  

    \foreach \x in {1,...,\numClasses}  
    {
    
    
    \begin{scope}[shift={(\x,-\y)}]
        \def\mVal{\myConfMat[\y-1][\x-1]} 
        \def\stdVal{\stdMat[\y-1][\x-1]}
        
        \def\r{\mVal}   %
        \def\rstd{\stdVal}   %
        \pgfmathtruncatemacro{\p}{round(\r/\totSamples*100)}
        \pgfmathtruncatemacro{\pstd}{round(\rstd/\totSamples*100)}
        
        \coordinate (C) at (0,0);
        \ifthenelse{\p<50}{\def\txtcol{black}}{\def\txtcol{white}} 
        \node[
            draw,                 
            text=\txtcol,         
            align=center,         
            fill=black!\p,        
            minimum size=\myScale*10mm,    
            inner sep=0,          
            ] (C) {$\p\pm\pstd\%$};     
            
        \ifthenelse{\y=\numClasses}{
        \node [] at ($(C)-(0,0.75)$) 
        {\pgfmathparse{\classNames[\x-1]}\pgfmathresult};}{}
    \end{scope}
    }
}
\coordinate (yaxis) at (-0.8,0.5-\numClasses/2);  
\coordinate (xaxis) at (0.5+\numClasses/2, -\numClasses-1.25); 
\node [vertical label] at (yaxis) {True grade};
\node []               at (xaxis) {Predicted grade};
\end{tikzpicture}
         }
        \label{fig:cm_grade}
    }~
    \sidesubfloat[]{
           \resizebox{0.45\textwidth}{!}{
\def\myConfMat{{
{1629.0,	20.6,	14.4},  
{12.4,	21.0,	0.6},  
{15.6,	2.4,	27.0},  
}}

\def\stdMat{{
{5.612486,	2.966479,	3.049590},  
{2.302173,	2.828427,	0.547723},  
{3.209361,	1.140175,	3.240370},  
}}

\def\classNames{{"normal","wedge","concave"}} 

\def\numClasses{3} 

\def\myScale{1.7} 
\begin{tikzpicture}[
    scale = \myScale,
    ]

\tikzset{vertical label/.style={rotate=90,anchor=east}}   
\tikzset{diagonal label/.style={rotate=45,anchor=north east}}

\foreach \y in {1,...,\numClasses} 
{
    \node [anchor=east] at (0.4,-\y) {\pgfmathparse{\classNames[\y-1]}\pgfmathresult}; 

    \def\totSamples{0}
    \foreach \ll in {1,...,\numClasses}
    {
        \pgfmathparse{\myConfMat[\y-1][\ll-1]}   
        \xdef\totSamples{\totSamples+\pgfmathresult} 
    }
    \pgfmathparse{\totSamples} \xdef\totSamples{\pgfmathresult}  

    \foreach \x in {1,...,\numClasses}  
    {
    
    
    \begin{scope}[shift={(\x,-\y)}]
        \def\mVal{\myConfMat[\y-1][\x-1]} 
        \def\stdVal{\stdMat[\y-1][\x-1]}
        
        \def\r{\mVal}   %
        \def\rstd{\stdVal}   %
        \pgfmathtruncatemacro{\p}{round(\r/\totSamples*100)}
        \pgfmathtruncatemacro{\pstd}{round(\rstd/\totSamples*100)}
        
        \coordinate (C) at (0,0);
        \ifthenelse{\p<50}{\def\txtcol{black}}{\def\txtcol{white}} 
        \node[
            draw,                 
            text=\txtcol,         
            align=center,         
            fill=black!\p,        
            minimum size=\myScale*10mm,    
            inner sep=0,          
            ] (C) {$\p\pm\pstd\%$};     
            
        \ifthenelse{\y=\numClasses}{
        \node [] at ($(C)-(0,0.75)$) 
        {\pgfmathparse{\classNames[\x-1]}\pgfmathresult};}{}
    \end{scope}
    }
}
\coordinate (yaxis) at (-0.8,0.5-\numClasses/2);  
\coordinate (xaxis) at (0.5+\numClasses/2, -\numClasses-1.25); 
\node [vertical label] at (yaxis) {True morphology};
\node []               at (xaxis) {Predicted morphology};
\end{tikzpicture}
         }
   \label{fig:cm_type}
  }%
        \caption{\textit{Confusion matrices.} Confusion matrices for the classification of each severity grade and morphology class, normalized by true classes.}
        \label{fig:cm}
\end{figure}

\newpage
\begin{table}[h!]
    \scriptsize
    \caption{\textit{Hyperparameters and training details.} Parameters used for training model components, chosen by grid search.}
    \label{tab:demographics}
    \centering
    {
    \renewcommand{\arraystretch}{1}
    \small
    \begin{tabular}{l|cc}
        \hline
         & Detector model & Keypoints model \\
         \hline
         \textit{learning rate} & $10^{-6}$ & $5\times 10^{-5}$ \\
         \textit{weight decay}  & $10^{-6}$ & $5\times 10^{-5}$ \\
         \textit{batch size}    & 8 & 64 \\
         \textit{equalization}  & \cmark & \cmark \\
         \textit{inversion}     & \cmark & \cmark \\
         \textit{rotation}      & 15\% & 20\% \\
         \textit{flips}         & \cmark & \cmark \\
         \textit{blur}          & \cmark & \cmark \\
         \textit{random bbox scale}  & \xmark & 35\%\\
         \textit{random bbox jitter} & \xmark & 10 pixels\\
         $\lambda_{iou}$ & 2 & - \\
         $\lambda_{\ell1}$ & 5 & - \\
         \hline

    \end{tabular}
    }

\end{table}



\begin{table}[h!]
    \centering
\caption{\textit{Ablation study of loss components.} 5-fold comparison of the vertebra morphology classification on the test set ($N_{vertebra}=\ntestvert$). \fscore, specificity and sensitivity computed at the Youden operating point. ($^\ast$) indicates significant difference to the full model using a Bonferroni $t$-test at $\alpha=0.05$.}
\label{tab:ablation_type}
{
\renewcommand{\arraystretch}{1}
\begin{tabular}{c|ccc|SSSS}
\hline
\multicolumn{1}{c}{OVR target} &
\multicolumn{1}{c}{$\L_{rle}$} &
\multicolumn{1}{c}{$\L_{kps}$} &
\multicolumn{1}{c}{$\L_{img}$} &

\multicolumn{1}{c}{AUC} &
\multicolumn{1}{c}{\fscore} &
\multicolumn{1}{c}{sensitivity} &
\multicolumn{1}{c}{specificity} 

\\
\hline

\multirow{4}{*}{\textit{normal}} & \checkmark &  &  
& \num{0.96 +- 0.01} 
& \num{0.48 +- 0.02}$^\ast$
& \num{0.90 +- 0.03}$^\ast$ 
& \num{0.89 +- 0.01}$^\ast$

\\
&  \checkmark &  \checkmark &   
& \num{0.96 +- 0.01} 
& \num{0.45 +- 0.06}$^\ast$  
& \num{0.91 +- 0.03}$^\ast$ 
& \num{0.86 +- 0.04}$^\ast$ 

\\
& \checkmark &  &  \checkmark  
& \bfseries\num{0.98 +- 0.01}$^\ast$
& \bfseries\num{0.71 +- 0.06}$^\ast$ 
& \num{0.91 +- 0.04} 
& \bfseries\num{ 0.96 +- 0.01} 

\\
& \checkmark &  \checkmark &  \checkmark  
& \num{0.92 +- 0.04} 
& \num{0.69 +- 0.08}  
& \bfseries\num{0.98 +- 0.01} 
& \num{0.95 +- 0.02} 

\\
 \hline
\multirow{4}{*}{\textit{wedge-like}} & \checkmark &  &   
& \num{0.96 +- 0.004}$^\ast$ 
& \num{0.95 +- 0.006}  
& \num{0.92 +- 0.012} 
& \num{ 0.93 +- 0.02} 

\\
& \checkmark &  \checkmark &   
& \num{0.95 +- 0.01}$^\ast$  
& \num{0.96 +- 0.001}
& \num{0.92 +- 0.002} 
& \num{0.90 +- 0.03} 

\\
& \checkmark &  &  \checkmark  
& \num{0.98 +- 0.017}
& \bfseries\num{0.98 +- 0.01} 
& \bfseries\num{0.96 +- 0.01}
& \num{0.92 +- 0.05} 

\\
& \checkmark &  \checkmark &  \checkmark  
& \bfseries\num{0.98 +- 0.01} 
& \num{0.97 +- 0.02} 
& \num{0.95 +- 0.04} 
& \bfseries\num{0.93 +- 0.04}

\\
\hline
\multirow{4}{*}{\textit{concave}} & \checkmark &  &   
& \num{0.94 +- 0.01}$^\ast$  
& \num{0.95 +- 0.01}$^\ast$ 
& \num{0.90 +- 0.03}$^\ast$  
& \num{0.86 +- 0.04}$^\ast$  

\\
& \checkmark &  \checkmark &   
& \num{0.93 +- 0.01}$^\ast$  
& \num{0.95 +- 0.01}$^\ast$   
& \num{0.92 +- 0.03} 
& \num{ 0.86 +- 0.03}$^\ast$ 

\\
& \checkmark &  &  \checkmark  
& \num{ 0.98 +- 0.01}
& \num{ 0.97 +- 0.02}  
& \num{0.95 +- 0.03} 
& \num{  0.925 +- 0.03}

\\
& \checkmark &  \checkmark &  \checkmark  
& \bfseries\num{0.98 +- 0.01} 
& \bfseries\num{0.98 +- 0.01}  
& \bfseries\num{0.96 +- 0.03}
& \bfseries\num{0.93 +- 0.03}

\\

\hline
\end{tabular}
}
\end{table}

\end{document}